\documentclass[a4paper,14pt]{extarticle}
\usepackage{tabularx}
\usepackage{longtable, float}
\usepackage{amssymb}
\usepackage{amsmath,bm}
\usepackage{amssymb}
\usepackage{graphicx}
\usepackage{amsfonts}         
\usepackage{fancybox}   
\usepackage{cite}

\usepackage[top=1.3cm, bottom=3.5cm, outer=2.5cm, inner=2.5cm, heightrounded,marginparwidth=1.5cm, marginparsep=0.6cm, margin=1.5cm]{geometry}

\usepackage[margin=1.2cm, font = small]{caption}

\usepackage[usenames,dvipsnames]{xcolor}
\usepackage[pagebackref, bookmarks={false}, pdfauthor={Homo Sapiens}, pdftitle={sykDef}]{hyperref}
\hypersetup{colorlinks=true, linkcolor=darkblue, citecolor=darkgreen, filecolor=OliveGreen, urlcolor=dgreen, filebordercolor={.8 .8 1}, urlbordercolor={.8 .8 0}} 
\usepackage{soul} 
\setstcolor{Red}

\usepackage{wrapfig}

\definecolor{dgreen}{rgb}{0.0, 0.42, 0.24}
\definecolor{darkgreen}{rgb}{0,0.4,0}
\definecolor{darkblue}{rgb}{0,0,0.4}

\def\sgn{{\rm sgn\ }}
\newcommand{\vev}[1]{\langle #1 \rangle}

\newcommand{\nn}{\nonumber}

\usepackage{subfigure}
\usepackage{mathtools}
\usepackage[english]{babel}
\usepackage[utf8x]{inputenc}
\usepackage[T1]{fontenc}

\renewcommand{\title}[1]{\vbox{\center\LARGE{#1}}\vspace{5mm}}
\renewcommand{\author}[1]{\vbox{\center#1}\vspace{5mm}}
\newcommand{\address}[1]{\vbox{\center\em#1}}

\renewcommand{\date}[1]{\vbox{\center#1}}

\begin{document}

\title{(Half) Wormholes under Irrelevant Deformation}

\author{Diptarka Das${}^{L}$, Sridip Pal${}^{R}$ and Anurag Sarkar${}^{L}$}

\address{ \vspace{0.4cm}
  {\it (L)
Department of Physics, 
 Indian Institute of Technology - Kanpur, \\
Kanpur 208016, India \\
 }
   {\it (R)
{School of Natural Sciences, Institute for Advanced Study,
Princeton, \\ NJ - 08540, USA.} }

 }


\abstract{ 
Recently it has been shown by Almheiri and Lin \cite{Almheiri:2021jwq} that the reconstruction of black-hole interior is sensitive to knowing the exact coupling of the boundary theory even if the coupling is irrelevant. This motivates us to enlarge the set of the one-time and two-time toy models inspired from the SYK by deforming the same with \textit{irrelevant} coupling. We find that both half-wormholes as well as the wormholes persist in presence of the deformation, leading to a similar mechanism for curing the factorization problem. While for the one time case, the deformed partition function and its moments change by an overall factor, which can completely be absorbed into a renormalization of coupling, for the two time (or coupled one time) SYK we find non-trivial dynamics of the saddles as the couplings are varied. Curiously, the \textit{irrelevant} deformations that we consider can also be thought of as an ensemble average over an overall scaling of the original undeformed Hamiltonian with an appropriate probability distribution, this allows for the possibility that half-wormholes may also be present in suitably defined ensemble of theories.

}


\vfill\eject


\section{Introduction and summary}

In AdS/CFT, the factorization problem arises if we have a bulk manifold $X$ with two (or multiple) disconnected conformal boundaries say $L$ and $R$. From the boundary perspective, the partition function $Z_{L\cup R}$ on $L \cup R$ should factorize as a product of two partition functions $Z_{L}$ and $Z_R$, defined solely on the $L$ and $R$ component respectively. On the other hand, the holographic dictionary instructs us to perform a bulk computation summing up contribution from all possible choice of X. This allows for in particular a contribution from wormhole geometries connecting the $L$ and $R$. Therefore the bulk computation seems to capture the correlation between two boundary quantities i.e $\langle Z_L Z_R\rangle-\langle Z_L\rangle\langle Z_R\rangle$, leading to non-factorization. Following \cite{Schlenker:2022dyo}, let us call these kind of contribution as \textit{connected amplitudes with disconnected boundaries} i.e CADB. They have recently gained lot of interest. In particular, the wormhole saddle contributions are responsible for the late time behaviours of the spectral form factor \cite{Saad:2018bqo,Saad:2019lba} and correlators \cite{Saad:2019pqd}, of holographic theories in the ramp regime, as well as the late time Hawking radiation entropy \cite{Penington:2019kki,Almheiri:2019qdq}. \\


The CADB leads to factorization problem only if one insists on having a unitary dual conformal field theory with fixed Hamiltonian. The puzzle can be avoided if one instead assumes the Hamiltonian is drawn from a random matrix ensemble. In fact, in 2D JT gravity, this is precisely what happens, the dual theory of JT gravity is not a single fixed theory rather an ensemble of theories \cite{Saad:2019lba}, hence provides a plausible explanation why CADB exists from the boundary perspective. Even though ensemble average saves a potential embarassment, this raises puzzle in cases where we expect and know the AdS/CFT duality to work without any averaging i.e when a theory with specific set of couplings in the boundary is dual to a bulk theory with specific set of parameters (as in the original examples of AdS/CFT based on maximally supersymmetric QFT models).\\

The dichotomy is that on one hand wormhole contributions are important for instance to find the ramp in the spectral form factor (SFF) while on the other hand they spoil factorization. Therefore it is desirable to search for wormholes even in the factorisable observables and naturally we are led to find other contributions which restore factorization in those scenarios. Moreover, the SFF is not self averaging \cite{PhysRevLett.78.2280}, for a single fixed theory there are erratic fluctuations around the ramp, hence there should be some bulk contribution responsible for these fluctuations. 
As the problem for the full SYK model is a difficult one, the authors of \cite{sssy, baur} focussed on the SYK model at a single time instance and within this toy problem resolved factorization by discovering \textit{half}-wormholes on top of the usual wormhole contributions. We emphasize that in these toy models, the wormholes survive even in the un-averaged theory and the half-wormholes added to it restore factorization. See also \cite{Saad:2021uzi} and \cite{Mukhametzhanov:2021hdi} on general mechanism of how half-wormholes can possibly cure the facorization puzzle. It worths mentioning that \cite{Blommaert:2021gha,Blommaert:2021fob} discuss $2$-dimensional gravity models where they can successfully address the factorization puzzle. While the former discusses an interpolating model between a random matrix ensemble and a fixed Hamiltonian, the latter utilizes non-local interactions in the action.\\

In this short note we enlarge the set of such toy examples by considering irrelevant deformations of the one (and two-) time SYK model.  Since the toy models are zero dimensional, any deformation is marginal. Our nomenclatures are made with reference to the actual SYK model. Let us motivate why we care about irrelevant deformations. First of all, one might wonder how robust is the prediction of one time SYK model in terms of having wormholes and half-wormholes. More importantly, it has been pointed out recently that the reconstruction of  black hole interior is highly theory sensitive \cite{ Almheiri:2021jwq}, in particular depends on the marginal and irrelevant couplings non-trivially. Hence it is a natural step to deform the original one time SYK model by irrelevant deformations and ask about the presence of wormholes and half-wormholes, to which we answer affirmatively. On top of the one time case, we also consider the coupled (two time instances) case which is closer to the actual SYK model. 

The relevant quantities that has been studied in \cite{sssy,baur} for the undeformed case are given by
\begin{equation}
\begin{aligned}
z_{\mathtt{SYK}} &= \int \mathrm{d}^N \psi \exp \left( H_0 \right),\\
\zeta_{\mathtt{SYK}}(\mu) &= \int \mathrm{d}^{2N}\psi\,\, \exp \left(H_{0L}+H_{0R}+ \mu \psi_i^L \psi^R_i \right)
\end{aligned}
\end{equation}
We note that $\zeta_{\mathtt{SYK}}(0)=z^2_{\mathtt{SYK}}$. Non-zero $\mu$ can be thought of as two-time or coupled one-time SYK model. One of the salient conclusions from \cite{sssy,baur} is 
\begin{equation}
z^2_{\mathtt{SYK}} \simeq \langle z^2_{\mathtt{SYK}}\rangle + \text{Half-Wormhole}
\end{equation}
and $\langle z^2_{\mathtt{SYK}}\rangle$ is identified with the wormhole contribution. Later we will refer them as moments of partition function. Furthermore, it has been shown in \cite{baur} that the wormholes appear as the final term in a perturbation series around the half-wormholes. Finally, the half-wormholes are responsible for curing the factorization problem.\\

We study the above physical observables for irrelevant deformation of $H_0$, denoted as $f(H_0)$. In particular, we expound upon
\begin{equation}
\begin{aligned}\label{eq:basic}
z_{\mathtt{deformed}} &= \int \mathrm{d}^N \psi \exp \left\{ f(H_0) \right\},\\
\zeta_{\mathtt{deformed}}(\mu) &= \int \mathrm{d}^{2N}\psi\,\, \exp \left\{f(H_{0L})+f(H_{0R})+ \mu \psi_i^L \psi^R_i \right\}
\end{aligned}
\end{equation}
 
 $\bullet$ Several interesting features come out of the analysis 
\begin{enumerate}
\item \textbf{(Half)-Wormholes persist after Deformation:} 

We find that both half-wormhole as well as the wormhole persist in presence of the deformation, leading to a similar mechanism for curing the factorization problem. 
\item \textbf{Wormholes as large fluctuations around Half-Wormhole even after Deformation:} 

Just like the original undeformed model, the wormhole can be thought of as a large fluctuation around the half-wormhole saddle. One need not add it separately. The scenario mimics the case of tensionless string where one can prove the background independence of the partition function\cite{Eberhardt:2021jvj}.
\item \textbf{Deformation $=$ Renormalization:} 

For the one time case, the deformed partition function changes by an overall factor, which can completely be absorbed into a renormalization of coupling $J$ amongst fermions. 
Quantitatively, consider the following deformation $f(H_0)$ ($E_0$ is the ground state energy of the undeformed model) and define the function $B$
\begin{equation}
f(H_0)=\sum_{\alpha=0}^{\infty}h(E_0,\alpha)H_0^{\alpha}\,,\quad \quad 
B(r_1)\equiv \sum_{\underset{\sum_i\alpha_i=p}{\{\alpha_i\geqslant 1\} }}\,\left(\prod_{i=1}^{r_1} h(E_0, \alpha_i)\right)\,,
\end{equation}
so that for the one time model, the renormalized coupling $J_{\mathtt{renormalized}}$ is given by
\begin{align}
J_{\mathtt{renormalized}}&= J \left(p! \sum_{r_1=1}^{p}\frac{1}{r_1!} B(r_1)\right)^{1/p}. \label{exact101}
\end{align}
For more details, see the end of \S\ref{review:syk}, in particular eq.~\eqref{exact1}.
For the two time (or coupled one time) SYK we find non-trivial dynamics of the saddles as the couplings are varied. In particular one can understand the deformation for the two-time case as renormalization of the two-time coupling $\mu$. In this case the wormholes lose their self-averaging characteristics if the renormalized $\mu_{\mathtt{renormalized}}$ becomes very large. On the other hand, in the regime, $1/N\ll\mu_{\mathtt{renormalized}}\ll1$, the wormholes persist and dominate over half-wormhole just like in the undeformed case.

\item \textbf{Deformation $=$ Ensemble average over Overall Scale:}

 The irrelevant deformations that we consider can also be thought of as an ensemble average over an overall scaling of the original undeformed Hamiltionian (along with possible shift in ground state energy) with an appropriate probability distribution. This allows for the possibility that the half wormhole may persist in a suitably defined ensemble theory as well. In particular, we can express eq.~\eqref{eq:basic} for appropriate $p(\ell)$, which depends on the deformation coupled with a possible shift in the ground state energy $E_0$:
 \begin{equation}
\begin{aligned}
z_{\mathtt{deformed}} &= \int\mathrm{d}\ell\ p(\ell) \int \mathrm{d}^N \psi \,\, e^{\ell (H_0-E_0) },\\
\zeta_{\mathtt{deformed}}(\mu) &=  \int\mathrm{d}\ell_L\mathrm{d}\ell_R\ p(\ell_L)p(\ell_R)\int \mathrm{d}^{2N}\psi\,\, e^{ \ell_L (H_{0L}-E_0)+\ell_R (H_{0R}-E_0)+ \mu \psi_i^L \psi^R_i }
\end{aligned}
\end{equation}

Specifically, for $T\bar{T}$ deformation and Gaussian deformation, we have
\begin{equation}
\begin{aligned}
p_{T\bar{T}\mathtt{deformed}}&=\frac{1}{\sqrt{8\pi\lambda \ell^3}} \exp \left\{ -\frac{(\ell-1)^{2}}{8 \lambda \ell}\right\} \quad \quad \ell \in [0,\infty)\,,\\
p_{\mathtt{Gaussian deformed}}&=\frac{1}{\sqrt{2\pi} s } \exp\left\{-\frac{(\ell-\ell_{*})^2}{2s_{}^2}\right\} \quad \quad \ell \in (-\infty,\infty)\,.
\end{aligned}
\end{equation}
Here $\lambda, s, \ell_*$ parametrize the deformation. For more details, see \S\ref{review:syk} and \S\ref{gdeform}.

It is worth mentioning that in \cite{Collier:2022emf}, it has been shown using harmonic analysis on $\mathrm{SL}(2,\mathbb{Z})$ that the fixed $N$ ensemble average of $\mathrm{SL}(2,\mathbb{Z})$ invariant physical observable over the $\mathcal{N}=4$ supersymmetric conformal manifold upon taking the large $N$ limit reproduces the large N, large `t-Hooft coupling limit. This provides another example where the ensemble average is equivalent to a theory with particular coupling, albeit in some particular limit. See also \cite{Benjamin:2021ygh} for the implication of harmonic analysis of $\mathrm{SL}(2,\mathbb{Z})$ in context of $2$D CFT.   In recent years, the ensemble average over some appropiate moduli space of $2$D CFT has been considered in many papers, which includes \cite{Maloney:2020nni,Afkhami-Jeddi:2020ezh,Datta:2021ftn,Benjamin:2021wzr,Dong:2021wot,Ashwinkumar:2021kav,Perez:2020klz,Dymarsky:2020qom,Dymarsky:2020pzc,Meruliya:2021lul,Meruliya:2021utr}, see also \cite{Cotler:2020hgz,Collier:2021rsn,Das:2021shw}. It deserves mention that the ensemble average over overall scaling induces a deformation is briefly discussed in the concluding section of \cite{Almheiri:2021jwq}.
\end{enumerate}

 
Deformations of the SYK theory have previously been studied in \cite{gross0, gross1}. The 1d analog of the $T\bar{T}$ deformation was shown to be equivalent to coupling an undeformed theory to the worldline quantum gravity. The generic deformation $H_0 \rightarrow f(H_0)$ is speculated to couple the undeformed theory to other 1d quantum gravities \cite{gross0}. 
For the SYK model, \cite{gross1} finds that the effect of the deformation $H(J) \rightarrow H(J) + \lambda T\bar{T}$, can essentially be captured in the renormalization of the coupling of the theory : $J \rightarrow J_{\mathtt{renormalized}} = J(\lambda)$. It is also quite important to normalize the vacuum energy by adding a shift: $ H_0 \rightarrow H_0 - E_0$. This constant shift though trivial for the undeformed theory, gives inequivalent deformed theories. In particular the renormalization becomes trivial when $E_0=0$. We show that similar features are also present in deformation of one and two-time SYK model. It deserves mentioning that the $T\bar{T}$ deformation in context of JT gravity is studied in \cite{Iliesiu:2020zld} (see also \cite{Stanford:2020qhm}).\\

The organization of the rest of the note along with glimpses of some results are as follows. In \S\ref{review:syk} we review the one-time undeformed SYK model, then go on studying its $T\bar{T}$ deformation. The deformed version is investigated in two different ways. The first method involves using einbein to recast the model as original SYK model with an intergral over the einbein variable. The integral over einbein for the $T\bar{T}$ deformation can also be interpreted as ensemble averaging. From the exact answer we conclude that both wormholes as well as half-wormholes are present in the deformed theory. The other method involves explicitly evaluating the $z^2_{\mathtt{deformed}}$ as a finite perturbation series around the half-wormhole saddle, where the wormhole appears as the final term in the series. We further generalize this second method to any arbitrary deformation $f(H_0)$. In \S\ref{two-time}, we look at two time SYK and its deformed version i.e study $\zeta_{\mathtt{deformed}}$ and dynamics of saddle points. As our work reveals interpretation of deformation in terms of ensemble averaging in \S\ref{gdeform} we study the Gaussian deformation. In particular, one chooses the overall scaling of the Hamiltonian from a Gaussian ensemble; this model is exactly equivalent to a quadratic deformation of the original Hamiltonian. Since this falls under the class of arbitrary deformation, the conclusion regarding the presence of (half)-wormholes remain true.


\section{The one-time SYK under $T\bar{T}$. }\label{review:syk}
In this section after a short review of the undeformed single-time SYK model, we analyse the deformed case. We write down the exact deformed partition function and its moments using both the einbein approach as well as through a Taylor expansion in the deformation parameter. We find that the deformed partition function is proportional to the undeformed single-time $z_\mathtt{SYK}$. Recalling that  $z_\mathtt{SYK}$ has half-wormholes and wormholes, we establish the presence of half-wormholes and wormholes in the deformed theory. The method involving the einbein further reveals that one can view the deformation as an ensemble average. Using the Taylor expansion method, one can conclude the presence of (half)-wormholes for $H\to f(H)$ for any nice enough function $f$.

\subsection{Undeformed theory} \label{hwh}
We start with the toy version of the undeformed SYK model wherein we focus on a single time instance. The analog of the partition function is built out of the following multi-Grassmann valued number : 
\begin{align}
H_0 &= i^{q/2} \sum_{1 \leq i_1 <\cdots <i_q\leq N} \,\,\, J_{i_1 \cdots i_q} \psi_{i_1 \cdots i_q} , \hspace{10mm} \psi_{i_1 \cdots i_q} \equiv \psi_{i_1}\psi_{i_2}\cdots\psi_{i_q}
\end{align}
The couplings $J_{i_1 \cdots i_q}$ are drawn randomly from a Gaussian distribution :
\begin{align} \label{j avg}
    \langle J_{i_1 \cdots i_q} \rangle =0, \hspace{10mm} \langle J_{i_1 \cdots i_q}J_{j_1 \cdots j_q} \rangle =\underbrace{\frac{(q-1)!}{N^{q-1}}}_{\bar{J}^{2}} \delta_{i_1 j_1}\cdots \delta_{i_q j_q}
\end{align}
We also assumed that both $N,q$ are even integers and the ratio $N/q = p$ is a positive integer. The partition function can be expressed as a Grassmannian integral:
\begin{align}
z_{\mathtt{SYK}} &= \int \mathrm{d}^N \psi \exp \left( H_0 \right) = \sum_{A_1 < \cdots < A_p} \sgn(A) J_{A_1} \cdots J_{A_p},
\end{align}
where in the last line the fermions were integrated out. The labels $A_i$ denote ordered non-intersecting subsets of $\{1, \dots , N \}$ with cardinality $q$. We remark that on non intersecting subsets $A_i=\{a_{i1},a_{i2},\cdots a_{iq}\}$ of $\{1, \dots , N \}$ with cardinality $q$ and $a_{im}<a_{in}$ if $m<n$, \textit{ordering} is defined as follows: $a_{i1}<a_{j1} \Leftrightarrow A_i<A_j$ for $i\neq j$. In the undeformed model with the explicit form of $z_{\mathtt{SYK}}$ we may evaluate the averaged (over $J$) moments  : $\vev{z^k_{\mathtt{SYK}}}$. Clearly, $\vev{z_{\mathtt{SYK}}} = 0$ since there are no possible Wick contractions of $J$ indices. The two lowest non-trivial moments comes from : 
\begin{align}
\vev{z^2_{\mathtt{SYK}} }&= \frac{N!}{p! (q!)^p } \left( \bar{J}^2 \right)^p, \label{z2z4-syk} \,\,\,\,
\vev{z^4_{\mathtt{SYK}}} = \left( \frac{\bar{J}^2 }{q!} \right)^{2p} \sum_{n_1 + n_2 +n_3 = p} \frac{ (q n_1)! (q n_2)! (q n_3)! }{( n_1!n_2! n_3!)^2}. 
\end{align}
In the large $N$ limit, it follows that 
\begin{align}
\vev{z^4 _{\mathtt{SYK}}} &\approx 3 \vev{z^2_{\mathtt{SYK}}}^2   \label{z4-syk}
\end{align}
 The correlations in the $\vev{z^2_{\mathtt{SYK}}} = \vev{z_L z_R}_{\mathtt{SYK}}$ contraction is the \emph{wormhole} contribution. We have added replica indices $L$ and $R$ to distinguish the replicas. For $\vev{z^4_{\mathtt{SYK}}}$ we therefore have : $\vev{z_L z_R z_{L'} z_{R'} }_{\mathtt{SYK}}$. Among all the possible contractions in the large $N$ limit the ones that dominate turn out to be: $\vev{z^4_{\mathtt{SYK}}} \approx \vev{z_L z_R }_{\mathtt{SYK}} \vev{z_{L'} z_{R'} }_{\mathtt{SYK}} + \vev{z_L z_{L'} }_{\mathtt{SYK}} \vev{z_{R} z_{R'} }_{\mathtt{SYK}}+ \vev{ z_L z_{R'} }_{\mathtt{SYK}} \vev{z_{R} z_{L'} }_{\mathtt{SYK}}$. It also turns out that each of the averages are equal to each other and equal to the wormhole correlation, from whence eq.~\eqref{z4-syk} follows. 
 
 \subsubsection*{(Half) wormholes of the undeformed theory}
The half-wormholes are the contributions to the unaveraged $z^2_{\mathtt{SYK}}$ in addition to $\vev{z^2_{\mathtt{SYK}}}$ which restore factorization. The regime where the half-wormholes dominate is best seen by introducing collective fields, $g \propto \sum_i \psi_i^L \psi_i^R$ and its conjugate $\sigma$. It turns out that the factorized answer can be expressed as: 
\begin{align}
z^2_{\mathtt{SYK}} &= \int \mathrm{d}\sigma\,\, \Phi_{0}(\sigma) \Psi_{0}(\sigma)\,,
\end{align}
where
 \begin{equation}
 \begin{aligned} \label{mean-00}
\Phi_0(\sigma) &=  \int \mathrm{d}^{2N} \psi \,\, \exp \bigg[ i e^{-i\pi/q} \sigma\psi_i^L \psi_i^R - \bar{J}^2\left( \psi_{A}^L \psi_{A}^R \right) + i^{q/2} J_A \left(\psi_A^L +  \psi_A^R \right) \bigg] \\
&=\int \mathrm{d}^{2N} \psi \,\, \exp \bigg[ i e^{-i\pi/q} \sigma\psi_i^L \psi_i^R - \frac{N}{q} \left( \frac{1}{N} \psi_i^L \psi_i^R \right)^q + i^{q/2} J_A \left(\psi_A^L +  \psi_A^R \right) \bigg]\,.
\end{aligned}
\end{equation}
Here we have used $\bar{J}^2=\frac{(q-1)!}{N^{q-1}}$ and $(\psi_{A}^L \psi_{A}^R)= \frac{1}{q!}(\psi_i^{L}\psi_i^{R})^q$ to go from the second line to the third line. The $\Psi_0(\sigma)$ is given by
\begin{equation}
\Psi_0(\sigma)=\int_{-\infty}^{\infty}\frac{\mathrm{d}g}{2\pi/N}\exp\left[N\left(-i\sigma g-\frac{1}{q}g^q\right)\right]\,,
\end{equation}
and is independent of the disorder.  Therefore disorder average, $\vev{\Phi_0(\sigma)}$ is well approximated by the contributions of the wormhole saddles. On the complex $\sigma$ plane these lie symmetrically on the unit circle separated out by angle $2\pi / q$. In order to compute the \texttt{RMS} we need 
\begin{equation}
\begin{aligned}
\vev{\Phi^2_0(\sigma)} &= \int \, \mathrm{d}^4g_{ab}\ \mathrm{d}^4\sigma_{ab}\ I(\sigma,\sigma_{ab},g_{ab})\,,
\end{aligned}
\end{equation}
where we have
\begin{equation}
\begin{aligned}
I(\sigma,\sigma_{ab},g_{ab})&\equiv\frac{1}{( 2\pi/N)^4} \exp\bigg[ N \log \bigg( \sigma^2+\sigma_{LR'}\sigma_{RL'}-\sigma_{LL'}\sigma_{RR'}\bigg) -N \bigg( i \sigma_{ab} g_{ab}  + \frac{g_{ab}^q}{q} \bigg)\bigg] . \label{rms2-00}
\end{aligned}
\end{equation}
where $(a,b)\in\{ (L,L'),(L,R'),(R,R'),(R,L') \}$ and $\sigma_{LR}=\sigma_{L'R'}$ fixed to $\sigma$ and not integrated over. $\sigma_{ab}=0$ is a trivial saddle for any value of $\sigma$ and responsible for $\vev{\Phi_0(\sigma)}^2$ contribution to $\vev{\Phi^2_0(\sigma)}$. When $\sigma=0$, the nontrivial saddles lie on unit circle and there are $q\times q\times 2$ of them. One of the two kinds corresponds to $(L,R'),(L',R)$ contraction while the other represents $(L,L'),(R,R')$ contraction. As we make $\sigma\neq 0$, these non trivial saddles take the form $\sigma_{LR'}\sigma_{RL'}=s^2e^{2\pi im/q}$, $\sigma_{LL'}\sigma_{RR'}=s^2e^{2\pi im/q}$. Without loss of generality, let us focus on the case $\sigma_{LR'}\sigma_{RL'}=s^2e^{2\pi im/q}$ and $\sigma_{LL'}\sigma_{RR'}=0$. Let us name the saddles as $\sigma_{ab*},g_{ab*}$. So we have two contributions : 
\begin{equation}
\begin{aligned}
\vev{\Phi^2_0(\sigma)}= I(\sigma,0,0) + \sum_{\text{Nontrivial Saddles}}I(\sigma, \sigma_{ab*},g_{ab*})\,.
\end{aligned}
\end{equation}
Now as we vary $\sigma$ on the complex plane, $\vev{\Phi^2_0(\sigma)}$ gets most of the contribution either from the trivial saddle or from the nontrivial ones. The region dominated by the trivial saddle is self averaging one while the other region is non self averaging.  Thus we can write
\begin{equation}
\begin{aligned}
\vev{\Phi^2_0(\sigma)}=\vev{\Phi_0(\sigma)}^2+ \sum_{\text{Nontrivial Saddles}}I(\sigma, \sigma_{ab*},g_{ab*})\,.
\end{aligned}
\end{equation}
%
 It turns out that for the one time SYK model, the wormhole saddles (which corresponds to some specific value of $\sigma$) always lie in the self-averaging region . The non-self averaging region takes the shape of a scallop (see Fig 1 in \cite{sssy} for $q=4$). For future reference, we define 
 \begin{align}
 \Theta(\sigma) &= \frac{\sum_{\text{Nontrivial Saddles}}I(\sigma, \sigma_{ab*},g_{ab*})}{\vev{\Phi_0(\sigma)}^2}\, , \label{Theta}
 \end{align}
 and use it as diagnostic of the self-averaging (or non self-averaging) region. The above exercise shows that the wormholes persist for the one-point SYK model before averaging since they live in the self-averaging region. 
 
Now we turn our attention to the half-wormholes. While the wormhole saddles are in self averaging regime, we know that $z^2$ is factorizable and can not be self-averaging exactly. This can be seen by setting $\sigma=0$ and noticing that the dominant contribution to $\vev{\Phi_0^2}$ come from non-trivial saddles, hence $\sigma=0$ is in non self-averaging region. It turns out that the following is a good approximation:
\begin{equation}
z^2_{\mathtt{SYK}}\simeq \langle z^2_{\mathtt{SYK}}\rangle+ \Phi_0(0)
\end{equation}
where $\Phi_0(0)$ is dubbed as \textit{half} wormhole \cite{sssy, baur} and responsible for factorization. The \texttt{RMS} value of the fluctuation is $\vev{\Phi^2_0(0)}$, which is dominated by nontrivial saddles is of the same order as the trivial contribution:
\begin{equation}
\sqrt{\vev{\Phi^2_0(0)}} \simeq  \vev{z^2_{\mathtt{SYK}}}
\end{equation}
Furthermore, one can show that other contributions to $z_\mathtt{SYK}^2$ are typically suppressed \cite{sssy, baur}.

\subsection{$T\bar{T}$ Deformed theory}
In this subsection we focus on the $\lambda T\bar{T}$ deformation. The deformation of the spectrum indexed by energy levels, $E$, can fully be solved from the non-perturbative flow equation: 
\begin{align}
\frac{\partial E}{\partial \lambda} &= \frac{E^2}{1/2 - 2 \lambda E}. 
\end{align}
The above can be integrated to find the full deformed Hamiltonian in terms of the original one as:
\begin{align}
H(\lambda) &= \frac{1}{4\lambda} \left( 1 \pm \sqrt{1 - 8\lambda( H_0 - E_0 ) }\right).
\label{TTb}
\end{align}
Where, $H_{0}$ is the undeformed SYK hamiltonian. We have also included a constant, $E_0$ shift to the ground state energy of the undeformed theory. The solution with negative sign is perturbatively connected from the undeformed spectrum, hence we consider, $H(\lambda)$ with only the negative sign in front of the square root. \\

In what follows we will compute the moments of this model using two methods. The first one involves introducing an einbein and then doing an ensemble average over the einbein to implement the deformation. The second method proceeds without the einbein by doing simple brute force Taylor expansion of deformed Hamiltonian around the original undeformed one. 



\subsubsection{Moments from exact computation-I via Einbeins} \label{sec:einbein}
A convenient way to implement the $T\bar{T}$ deformation given by eq.~\eqref{TTb} is to use an intergral over einbeins. The zeroth moment i.e partition function now involves also an integral over the einbein $\ell$ as follows\footnote{We thank Baur Mukhametzhanov for pointing out the one loop exactness of the deformation.}:

\begin{align}
z_{\mathtt{deformed}} &= \int_0^\infty \mathrm{d}\ell\ \frac{1}{\sqrt{8\pi\lambda \ell^3}} \exp \left\{ -\frac{(\ell-1)^{2}}{8 \lambda \ell} \right\} \int \mathrm{d}^{N}\psi \, \exp \left\{ \ell (H_{0} -E_0) \right\} \label{z einbein}. 
\end{align}
We note that $\frac{1}{\sqrt{2\pi\lambda \ell^3}} \exp \left\{ -\frac{(\ell-1)^{2}}{8 \lambda \ell} \right\}$ can be thought of as a probability distribution of $\ell$, the overall scaling of shifted Hamiltonian $H_0-E_0$. The distribution has its support on $[0,\infty)$ with unit mean and $4\lambda$ variance. This is only when the deformation coupling $\lambda$ is positive. \footnote{ When $\lambda < 0$, the analog process with the probability distribution $\frac{1}{\sqrt{2\pi\lambda \ell^3}} \exp \left\{ -\frac{(\ell+1)^{2}}{8 \lambda \ell} \right\}$ having support on $\ell \in (-\infty, 0]$ with mean $-1$ and variance $-4\lambda$, yields, $e^{-H(\lambda)}$. } As promised in the introduction, we see the ensemble average over overall scaling implements a deformation of the original Hamiltonian. \\

We can rescale $\psi$ in $H_0$ to separate out the $\ell$ integral so that we obtain
\begin{align}
&z_{\mathtt{deformed}} = \int \mathrm{d}\ell\  e^{-S'(\ell)} \, \int \mathrm{d}^{N}\psi \, \, \exp \left\{   i^{q/2} \sum_{1 \leq i_1 <\cdots <i_q\leq N} \,   J_{i_1 \cdots i_q} \psi^L_{i_1 \cdots i_q} \right\}\\
&\text{with, } \, S'(\ell) = \frac{ (\ell - 1)^2}{ 8 \lambda \ell}  + \ell E_0+\frac{1}{2}\log(8\pi\lambda)+\left(\frac{3}{2}-\frac{N}{q}\right)\log\ell.\nn 
\end{align}

Note that we have used the Grassmannian identity $d ( a \psi ) = d \psi / a$ above. The $\ell$ integral can be now be done explicitly to obtain
\begin{equation}
z_{\mathtt{deformed}}=\left( \frac{e^{\frac{1}{4 \lambda }}\left(1+8 \lambda  E_0\right)^{\frac{1}{4}-\frac{N}{2 q}}  }{\sqrt{2 \pi\lambda }}  K_{\frac{1}{2}-\frac{N}{q}}\left[\frac{\sqrt{1+8 \lambda  E_0}}{4 \lambda }\right]\right)z_{\mathtt{SYK}} \label{eq:besselexact}
\end{equation}

Since the effect of the deformation appears as an overall factor to $z_\mathtt{SYK}$ the existence of wormholes and half-wormholes follow naturally. One can also check this by looking at the variance and moments, e.g., the $k$ th moment of the partition function upon performing the SYK like average over the coupling $J$ evaluates to:

%
%
%

\begin{equation}
\vev{z^k_{\mathtt{deformed}}}\equiv  \int \left(\prod_A \frac{\mathrm{d}J_A}{\sqrt{2\pi}\bar{J}}\right) e^{-\frac{1}{2\bar{J}^2}J_A^2} z^k_{\mathtt{deformed}}(J)
\end{equation}
and noting the integral only affects the $z^k_{\mathtt{SYk}}$ factor of $z^k_{\mathtt{deformed}}$. Thus
\begin{equation}
\begin{aligned} 
\vev{z^k_{\mathtt{deformed}}} &=\left( \frac{e^{\frac{1}{4 \lambda }}\left(1+8 \lambda  E_0\right)^{\frac{1}{4}-\frac{N}{2 q}}  }{\sqrt{2 \pi\lambda }}  K_{\frac{1}{2}-\frac{N}{q}}\left[\frac{\sqrt{1+8 \lambda  E_0}}{4 \lambda }\right]\right)^k \vev{z^k_{\mathtt{SYK}}} \\
\end{aligned}
\end{equation}

Note that in the limit $\lambda\to 0$ with $\lambda E_0$ held fixed, we can use the asymptotics of Bessel $K$ to obtain
\begin{align}
\vev{z^2_{\mathtt{deformed}}} &\approx \frac{e^{2E_0'}}{\left( 1 + 8 \lambda E_0 \right)^{p} } \vev{z^2_{\mathtt{SYK}}}, \,\,\, E_0' = \frac{1- \sqrt{1 + 8 E_0 \lambda }}{4\lambda}. \label{z2av-onetime}
\end{align}
Similarly, we have
\begin{align}
\vev{z^k_{\mathtt{deformed}}}&\approx \frac{e^{k E_0'}}{\left( 1 + 8 \lambda E_0 \right)^{pk/2} }\vev{z^k_{\mathtt{SYK}}}
\end{align}

The same conclusion will 
 be reached using the exact computation without using einbein. 


\subsubsection{Moments from exact computation-II}
In this subsection we evaluate the Grassmanian integrals exactly to compute one-time $z$ and its higher moments. We find that in the leading order in $N$, and more specifically when $N \lambda \ll q$, the answers from the einbeins are reproduced. Our methods allow for a easy generalization to arbitrary deformations of the type $H_0 \rightarrow f(H_0)$ of which $T\bar{T}$ is only a special case. For the latter we can write: 
\begin{equation}
\begin{aligned} 
z_{\mathtt{deformed}}  = \int \mathrm{d}^N \psi \, e^{H(\lambda)}& = \int \mathrm{d}^N \psi \, \exp \left\{ \frac{1}{4\lambda} \left( 1-\sqrt{1-8 \lambda (H_0-E_0) } \right) \right\} \\ 
&= \int \mathrm{d}^N \psi \, \sum_{r=0}^{\infty} \, \frac{1}{r!} \left( \sum_{\alpha=0}^{\infty} \, h(E_0, \alpha) \, H_{0}^{\alpha}  \right)^{r}\,,
\end{aligned}
\end{equation}
where $h(E_0,\alpha)$ is given by
\begin{equation}
\begin{aligned}
h(E_0, \alpha)&=\left( \sum_{k \geq \alpha\,k>0}^{\infty} \, \frac{1}{4\lambda E_0} \binom{1/2}{k}\binom{k}{\alpha} \, (8 \lambda E_0)^{k} \, (-E_{0})^{1-\alpha} \right)\\
&=\begin{cases}2^{3 \alpha -2} \binom{1/2}{\alpha } \left(\frac{-\lambda}{1+8\lambda E_0}\right)^{\alpha} \frac{\sqrt{1+8\lambda E_0}}{-\lambda}\,&\text{for}\,\ \alpha\geq 1\,,\\
\frac{1}{4\lambda}\left(1-\sqrt{1+8\lambda E_0}\right)\,&\text{for}\,\ \alpha=0\,.
\end{cases}
\end{aligned}
\end{equation}
Now let us consider $\left( \sum_{\alpha=0}^{\infty} \, h(E_0, \alpha) \, H_{0}^{\alpha}  \right)^{r}$ and separate out the $\alpha=0$ term from the rest, so that we have
\begin{equation}
\left(  h(E_0, 0) + \sum_{\alpha=1}^{\infty}  h(E_0, \alpha)  H_{0}^{\alpha}  \right)^{r}=\sum_{r_1=0}^r \binom{r}{r_1} h(E_0, 0)^{r-r_1}\left(\sum_{\alpha=1}^{\infty}  h(E_0, \alpha)  H_{0}^{\alpha}\right)^{r_1}\label{expansion}
\end{equation}
Now the idea is that we need exactly $N$ Grassmannian variables to saturate the integral, thus we need to figure out the coefficient of $H^{p}$, where $N/q=p$.
From eq.~\eqref{expansion} we find that:
\begin{equation}\label{def:B}
B(r_1)\equiv \sum_{\underset{\sum_i\alpha_i=p}{\{\alpha_i\geqslant 1\} }}\,\left(\prod_{i=1}^{r_1} h(E_0, \alpha_i)\right)\,.
\end{equation}
The sum runs over all possible sets $\{\alpha_i: i=1,2,\cdots r_1\}$ such that $\sum_i\alpha_i=p$ and $\alpha_i>0$. The partitions are unordered, e.g., in the length $r_1 = 3$ partition of $p = 4 = 1+1+2$, the combination contributes with degeneracy $3$ as $(1,1,2), (1,2,1)$ and $(2,1,1)$. Also note that $$B(r_1)=0\,,\text{if}\, r_1>p\,.$$
This happens because $\alpha_i\geqslant 1$ and hence $p=\sum_i\alpha_i \geqslant r_1$.
Finally performing the Grassmanian integral, we find that
\begin{equation}\label{eq:deform}
z_{\mathtt{deformed}} =e^{h(E_0, 0) } \left(\sum_{A} \mathrm{sgn}(A) J_{A_1}J_{A_2}\cdots J_{A_p}\right)\left(p! \sum_{r_1=1}^{p}\frac{1}{r_1!} B(r_1)\right)
\end{equation}
One can check numerically that the above equation is identical to eq. \eqref{eq:besselexact}, for more details see Appendix \S \ref{app:bessel}. $B(r_1)$ is maximized when $r_1=p$ and the most contributing term for small $\lambda$ is coming from the set $\alpha_i=1$. The next leading term comes from $B(p-1)$, where the most contributing term is when one of the $\alpha_i=2$ and rest are set to $1$:
\begin{equation}
\begin{aligned}
B(p)&\simeq \left(1+8\lambda E_0\right)^{-p/2}\\
B(p-1)&\simeq 2(p-1)\lambda\left(1+8\lambda E_0\right)^{-p/2-1}
\end{aligned}
\end{equation}
Hence in the approximation $p\lambda \ll 1$, we have
\begin{equation}
z_{\mathtt{deformed}} =\frac{1}{\left(1+8\lambda E_0\right)^{p/2}} \exp\left[\frac{1-\sqrt{1+ 8 E_0 \lambda}}{4\lambda}\right]\left(\sum_{A} \mathrm{sgn}(A) J_{A_1}J_{A_2}\cdots J_{A_p}\right)+ O\left(\frac{p\lambda}{1+8\lambda E_0}\right). \label{z-exact}
\end{equation}
If we now consider $\vev{z^2_{\mathtt{deformed}} }$ at the leading order it will precisely give us eq.~\eqref{z2av-onetime}.
Going back to eq.~\eqref{z-exact} we can use the replacements : \begin{equation} J \rightarrow J(\lambda) =\frac{1}{\sqrt{1+8\lambda E_0}}J\end{equation} and $E'_0$ from eq.~\eqref{z2av-onetime} to rewrite:
\begin{equation}
z= e^{E'_0}\left(\sum_{A} \mathrm{sgn}(A)J_{A_1}(\lambda)\cdots J_{A_p}(\lambda)\right) + O\left(\frac{p\lambda}{1+(4\lambda E'_0)^2}\right)
\end{equation}
In leading order, this is therefore the one time SYK model with renormalized coupling $J(\lambda)$ and ground state energy shifted to $E'_0$. This is consistent with the results obtained in the full SYK model using diagrammatics and einbeins \cite{gross1}.

\subsubsection*{Arbitrary deformations}\label{sec:def}
For arbitrary Hamiltonian deformations  $H_0 \rightarrow f(H_0)$, we therefore see that once we know the Taylor coefficients $f(H_0) = \sum_\alpha h(E_0 , \alpha) H_0^\alpha$, the partition function reduces to $z_{\mathtt{SYK}}$ with renormalized couplings, 
\begin{align}
J_{A}{}_{\mathtt{renormalized}} &= J_A \left(p! \sum_{r_1=1}^{p}\frac{1}{r_1!} B(r_1)\right)^{1/p}. \label{exact1}
\end{align}
and ground state shifted $E_0 \rightarrow h(E_0,0)$, where $B(r_1)$ is defined in eq.~\eqref{def:B}.


\section{Two time SYK and deformation}\label{two-time}
\begin{figure}[tb]
 \centering
  \includegraphics[scale=0.7]{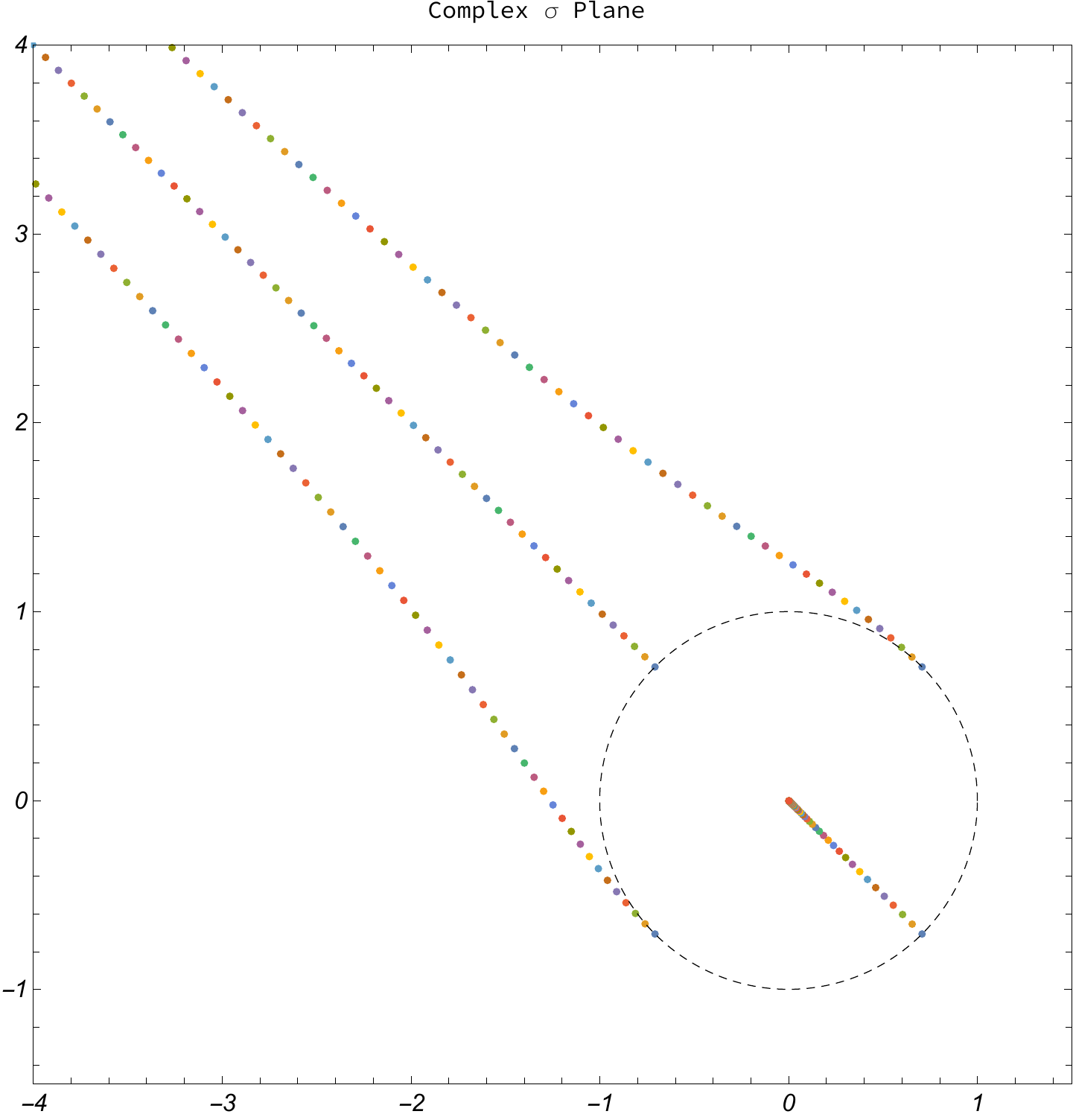} 
  \caption{The saddles contributing to $\vev{\zeta(\mu) }$, is plotted for $q=4$. The saddles start out from the four rotated roots of unity. Different colors indicate different values of $\mu$. As $\mu$ increases while one saddle goes to zero, the other three merge at $\mu e^{i3\pi/4}$. }
  \label{fig:sad}
 \end{figure}

In this section we consider the two time version of the SYK model and its deformation. We begin with undefomred theory, which can be thought of as coupling two one time models with coupling denoted as $\mu$. 
\begin{align}
\zeta(\mu) &= \int d^{2N}\psi\,\, \exp \left\{ i^{q/2} J_A \left(  \psi_A^L +  \psi_A^R \right) + \mu \psi_i^L \psi^R_i \right\}
\end{align}
Note, $\zeta(0) = z_{\mathtt{SYK}}^2$ and we have
\begin{equation}
\begin{aligned}
\zeta(\mu)&= \int \mathrm{d}\sigma\,\, \Phi_{0}(\sigma-i e^{i\pi/q}\mu ) \Psi_{0}(\sigma)\,,
\end{aligned}
\end{equation}
where we repeat
\begin{equation}
\Psi_0(\sigma)=\int_{-\infty}^{\infty}\frac{\mathrm{d}g}{2\pi/N}\exp\left[N\left(-i\sigma g-\frac{1}{q}g^q\right)\right]\,,
\end{equation}
Now averaging leads to 
\begin{equation}
\begin{aligned}
\vev{\zeta(\mu)}&= \int \mathrm{d}\sigma\,\, \left(i e^{-i\pi/q}\sigma+\mu\right)^N\int_{-\infty}^{\infty}\frac{\mathrm{d}g}{2\pi/N}\exp\left[N\left(-i\sigma g-\frac{1}{q}g^q\right)\right]\,,
\end{aligned}
\end{equation}
At large $N$ the above integrals over $\sigma$ and $g$ maybe computed by saddle point. If we solve the saddle point equations simultaneously we can get the saddle points as specific values of $\sigma$. These are the analogs of the wormhole contributions since a non-zero saddle point value of $\sigma$ implies a non-vanishing correlation between the two replicas. Unlike the one time case, for a non-zero $\mu$ the wormhole saddles are no longer present on the unit circle in the complex $\sigma$ plane, though there are still $q$ of them. We plot in Fig.\ref{fig:sad} for the $q=4$ case how the saddles behave as a function of $\mu$. When $\mu= 0$ the saddles appear symmetrically on the unit circle as it should since $\zeta(0) = z_{\mathtt{SYK}}^2$. As $\mu$ gets turned on, the saddles start to move asymmetrically.  One of the saddles always approach the origin $\sigma=0$. As $\mu$ increases, the rest $q-1$ saddles collapse and 
start moving at an angle $\tfrac{\pi}{2} + \tfrac{\pi}{q}$. 

 \subsubsection*{(Half) wormholes of the undeformed (two-time) theory}\label{hwhwithmu}
We repeat the exercise done in the end of \ref{hwh}, but now with $\Phi_{0}(\sigma-i e^{i\pi/q}\mu )$.
 When $\mu=0$ the self averaging region takes the form of a scallop centered at origin. The wormhole saddles live outside the scallop and hence also exist in the non-averaged model. We contrast this with the large $\mu$ where the scallop is now centered at $\sigma=ie^{i\pi/q}\mu$ i.e.\! the center moves along a line with an angle $\pi/2+\pi/q$. For large enough $\mu$ the $\sigma\neq 0$ saddles comes within the scallop. Thus for large enough $\mu$, the non averaged model does not have the wormhole saddles. This is expected since increasing $\mu$ effectively makes the model a $q=2$ non-averaged model, which is a Gaussian integrable theory. Effects of disorder in the spectral form factor for the $q=2$ SYK model have been analyzed in \cite{Liao:2020lac, Winer:2020mdc}, wherein the ramp is exponential. On the other hand we know that wormholes give rise to a linear ramp.
 \\
 
In Fig.\ref{fig:self-sad} we illustrate the dissolution process of the wormhole saddles with increasing $|\mu|$.  The plots take into account both the movement of the wormhole saddles as well as the motion of the scallop.  A simple dimension analysis shows that  the threshold value $\mu_*$ scales with variance in following manner :
\begin{align}
\mu_* \propto \left(\bar{J}^{2}\right)^{1/q}. 
\end{align}
Also as we see in the appendix \S\ref{app:scallop}, there is a $q$ dependence on the size of the scallop. With increasing $q$ the scallop gets smaller, and therefore the saddles survive further out in the complex $\sigma$ plane as compared with a lower $q$ result. Intuitively as increasing $q$ increases; chaoticity this is also what one would expect.

 \subsubsection*{(Half) wormholes of the deformed (two-time) theory}\label{hwhwithmuandDeform}
At this point, it is easy to figure out the effect of deformation. We will be brief and mention the salient features only. 
\begin{align}
\zeta_\mathtt{deformed}(\mu) &= \int d^2 \ell \,\,e^{-S(\ell_L, \ell_R)} \int d^{2N}\psi\,\, \exp \left\{ i^{q/2} J_A \left( \ell_L \psi_A^L + \ell_R \psi_A^R \right) + \mu \psi_i^L \psi^R_i \right\} \nn \\
&= \int \mathrm{d}\sigma\,\, \Phi_\mathtt{deformed}(\sigma-i e^{i\pi/q}\mu) \Psi_0(\sigma)\,,
\end{align}
where  $\Phi_\mathtt{deformed}(\sigma-i e^{i\pi/q}\mu) = \Phi_0(\sigma-i e^{i\pi/q}\mu_\mathtt{renormalized})$, which can be obtained by carrying out the einbein integrals via saddle. Thus we can also express $\zeta_\mathtt{deformed}(\mu) = \zeta(\mu_\mathtt{renormalized})$.
Of course note, $\zeta(0) = z^2$. 
For the $T\bar{T}$, at leading order in $p\lambda \ll 1$ we find that the dissolution of the wormhole saddles now happen for a different value of $\mu$ :
\begin{align}
\mu_{*\mathtt{renormalized}} &\propto \left( \bar{J}^2_\mathtt{renormalized}\right)^{1/q} = \frac{\left(\bar{J}^2\right)^{1/q} } { \left( 1 + 8 E_0 \lambda \right)^{1/q} }\nn \\
&= \mu_* \left( 1 + 8 E_0 \lambda \right)^{-1/q}.
\end{align}
We therefore see for $\lambda<0$ (recall $E_0$ is negative), if we increase $|\lambda| $, the renormalised variance $\bar{J}_\mathtt{renormalized}$ decreases and therefore $ \mu_{*\mathtt{renormalized}} < \mu_*$, hence the wormholes dissolve faster. 

\begin{figure}[!ht]
\centering

\includegraphics[scale=0.5]{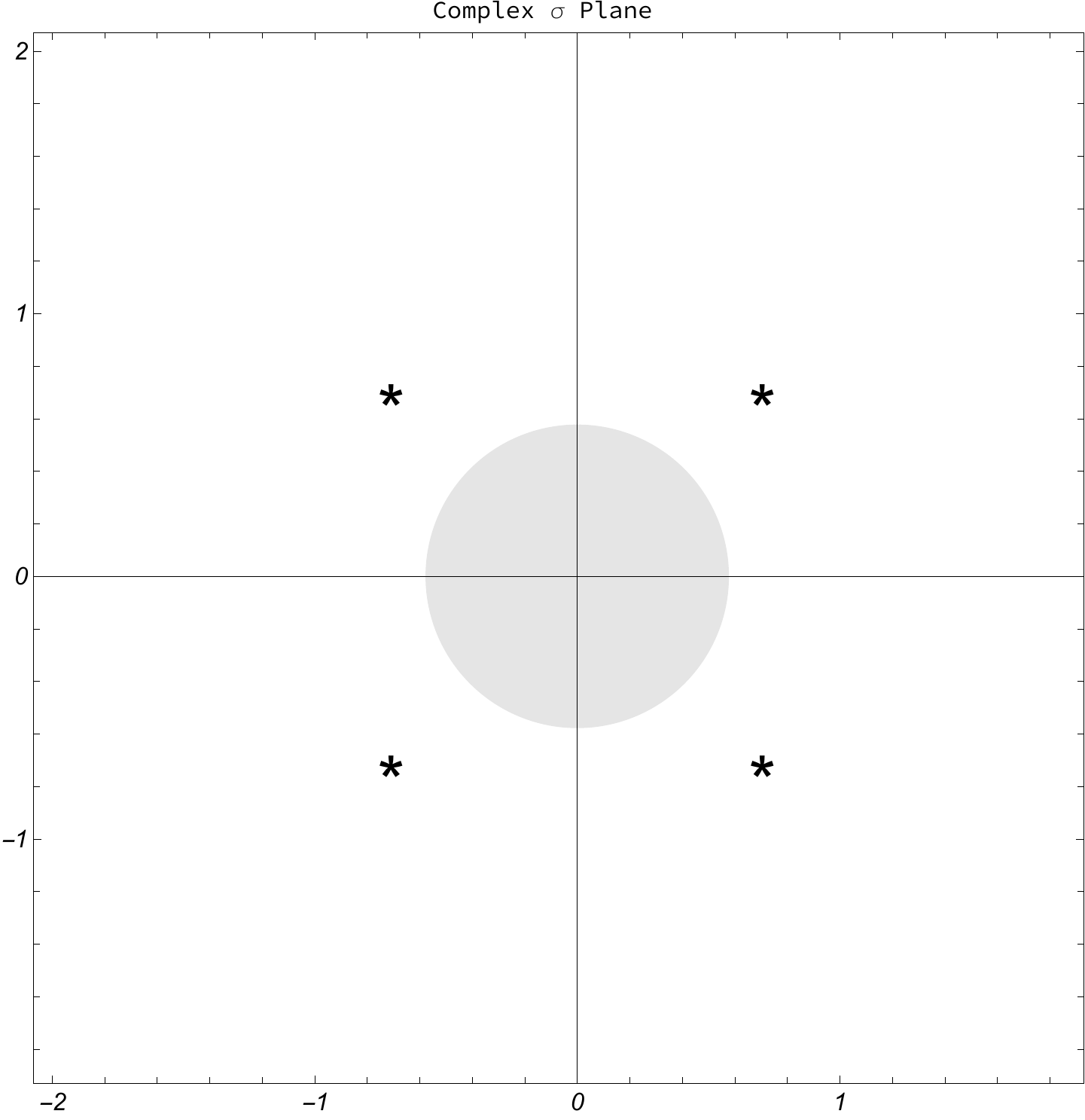}
\caption{The plots are for the $q=4$ case. The gray regions indicate approximately the scallop which is \emph{not} self-averaging in the complex 
$\sigma$ plane. The black stars indicate the wormhole saddles. For $\mu=0$, all the saddles are in non-self averaging region.}
\label{fig:zero}
\end{figure}

\begin{figure}[!ht]
\centering
\subfigure[]{
\includegraphics[scale=0.5]{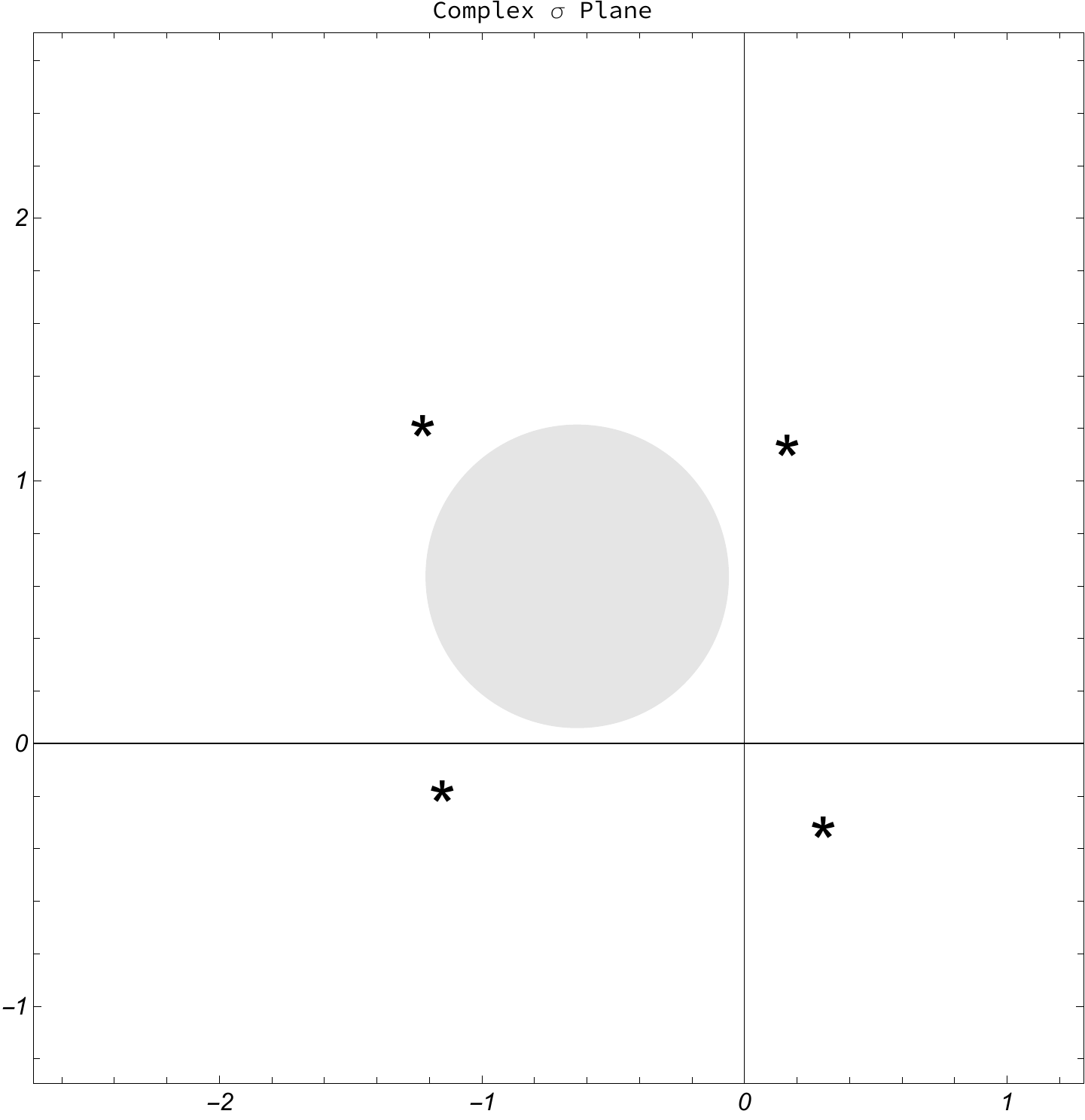}
\label{fig:one}
}
\subfigure[]{
\includegraphics[scale=0.5]{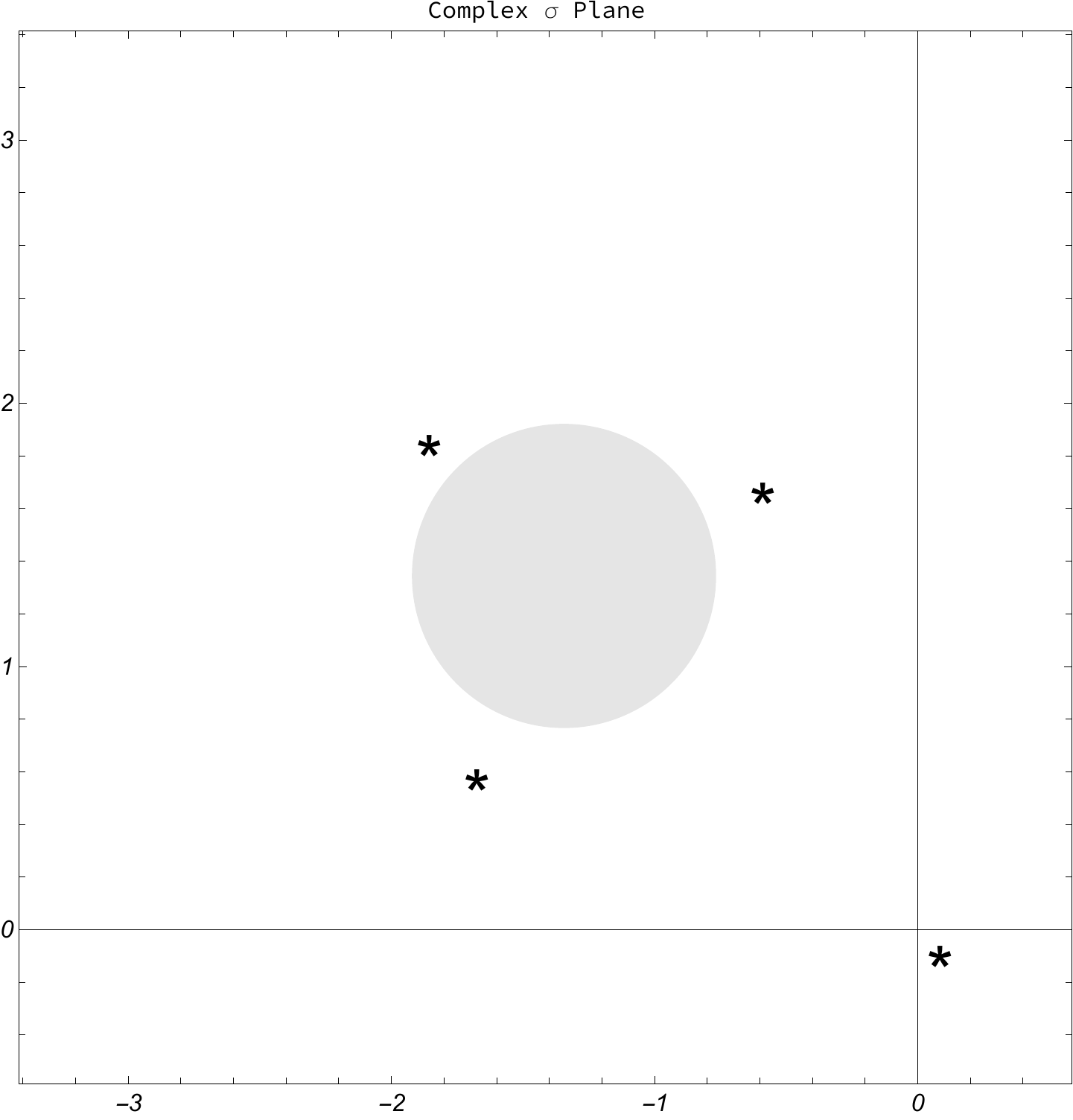}
\label{fig:two}
}
\subfigure[]{
\includegraphics[scale=0.5]{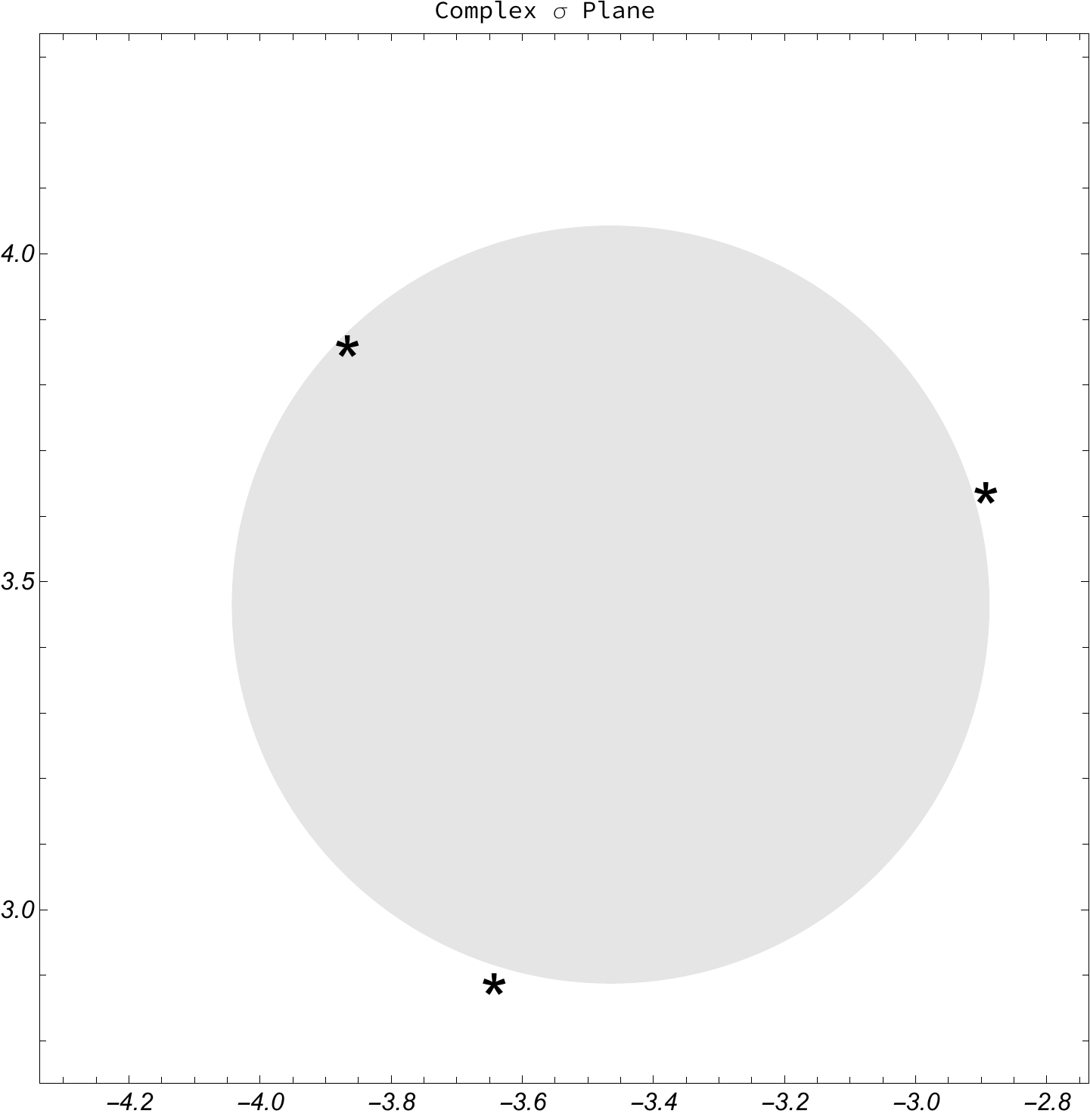}
\label{fig:two}
}
\subfigure[]{
\includegraphics[scale=0.5]{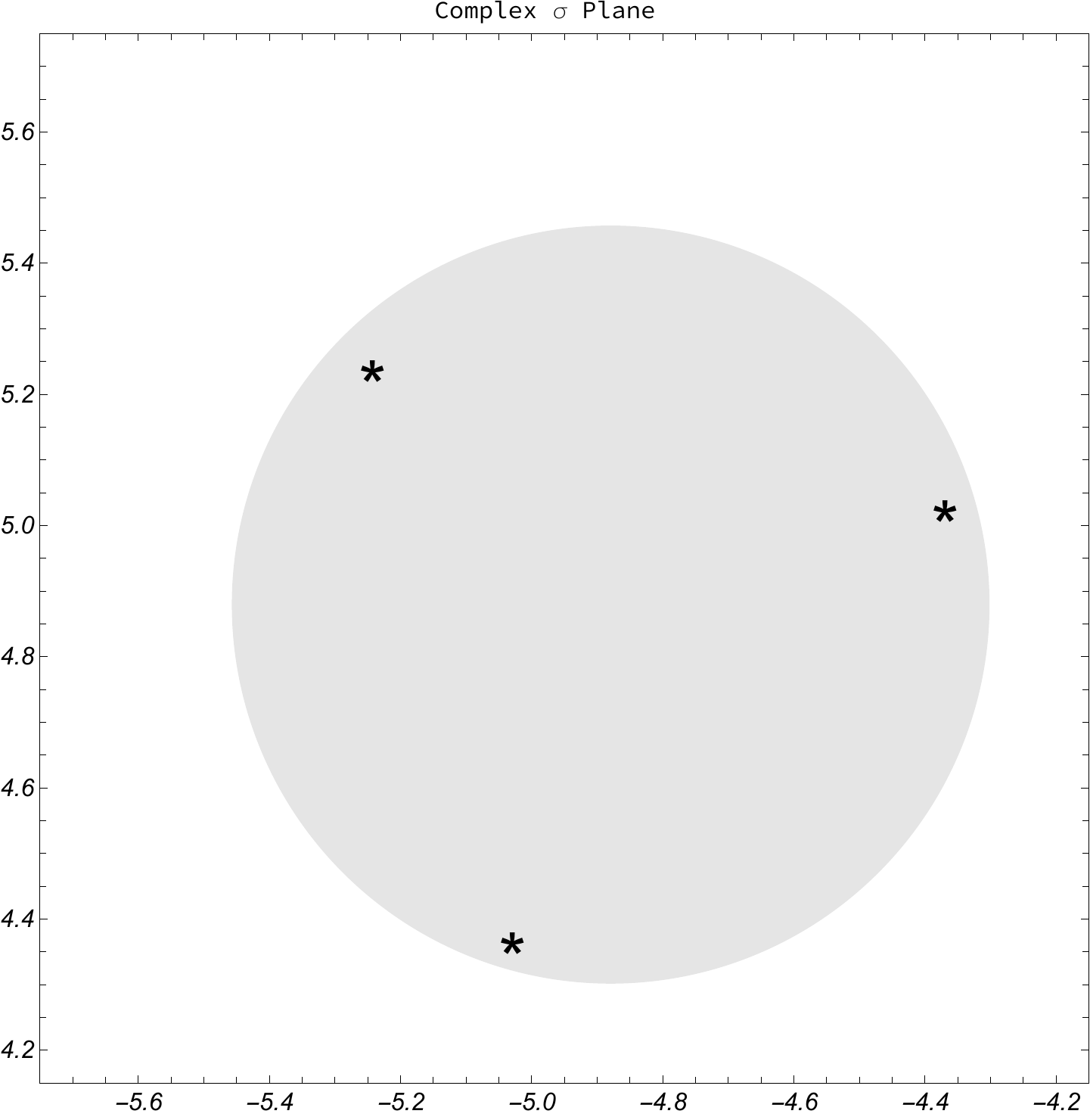}
\label{fig:three}
}
\caption{The plots are for the $q=4$ case. The gray regions indicate approximately the scallop which is \emph{not} self-averaging in the complex 
$\sigma$ plane. The black stars indicate the wormhole saddles. Progressive plots are for increasing values of $\mu$. The figure focuses on the three saddles which are engulfed, while the other one (not shown in the last two) remains very close to zero.}
\label{fig:self-sad}
\end{figure}

Since the $1/N$ expansion in this toy model is finite it is possible to expand $\zeta_\mathtt{deformed}$ also around the half-wormhole i.e. $\sigma= 0$ saddle \cite{baur} : 
\begin{align}
\zeta_\mathtt{deformed} &= \sum_{k=0}^p \zeta_k (\mu_\mathtt{renormalized}) \tilde\Phi_k, \label{phik}
\end{align}
where, 
\begin{align}
\tilde\Phi_k &= \frac{1}{k!} \left( \bar{J}^2 \right)^k \left( - \frac{\partial}{\partial \bar{J}^2 } \right)^k \Phi_{\mathtt{deformed}}(0), \,\,\,\,\,\\ \zeta_k(\mu_\mathtt{renormalized}) &= \sum_{n=0}^k \frac{k!}{( k-n)! ( n q)! } \left( \frac{ q}{N} \right)^n \left( N \mu_\mathtt{renormalized} \right)^{nq}. 
\end{align}
In terms of this expansion the wormhole saddles appear as the final terms in the series. Also note that in this expansion the effect of the deformation is encapsulated via $\mu \rightarrow \mu_\mathtt{renormalized}$. In the next section we shall look into this expansion for the exactly solvable case of Gaussian ensemble of einbeins.


\section{The Gaussian deformation}\label{gdeform}
One of the key objects that we have investigated so far is: 
\begin{align}
\zeta_\mathtt{deformed} &= \int \mathrm{d}^2 \ell \,\,e^{-S(\ell_L, \ell_R)} \int \mathrm{d}^{2N}\psi\,\, \exp \left\{ i^{q/2} J_A \left( \ell_L \psi_A^L + \ell_R \psi_A^R \right) + \mu \psi_i^L \psi^R_i \right\}
\end{align}
The purpose of the einbein integral is to implement the irrelevant deformation $H \rightarrow f(H)$. One can view the above as an ensemble average with a weight factor of $e^{-S(\ell_L, \ell_R)}$. Since $S(\{\ell_i\}) = \prod_i S(\ell_i)$ this averaging does not induce correlations between the {\em replicas}  for $\mu=0$, rather just renormalizes $J_A$ which is now a coupling of the irrelevant Hamiltonian, $f(H)$. On the other hand for $\mu\neq0$ , things are non trivial. One can not simply scale out $\ell$, and cast the deformation as an overall factor. \\

The choice of $S(\ell_L,\ell_R)$ provides a class of toy models where one can replace $e^{-S(\ell_L, \ell_R)}$ with a probability measure. An exactly solvable choice is the Gaussian ensemble for the einbeins $\ell_i$ with means $\ell_i^*$ and variances $s_i$. For simplicity, we will assume $\ell_{L*}=\ell_{R*}=\ell_*$. It results in the exact deformation: 
$
H_0 \rightarrow \frac{1}{2} (  s^2 H_0^2 + 2 \ell_* H_0 ).
$
In general for the deformation of the type : $ H_0 \rightarrow \# H_0^\gamma$, the einbein ensemble action that implements this (in the saddle-point approximation) is $S(\ell) = \# \ell^{\gamma/(\gamma-1)}$. Coming back to the Gaussian deformation, the one-time partition function is given by:
\begin{align}
z_{\mathtt{deformed}} &= \vev{\ell^p}_s z_\mathtt{SYK},  \label{1times} \\
\text{where, } \langle \ell^{k}\rangle_{s}&= \int_{-\infty}^\infty \mathrm{d}\ell\ \ell^k e^{-S(\ell)} =\frac{1}{\sqrt{2\pi} s } \int_{-\infty}^\infty  \mathrm{d}\ell \,\, \ell^{k} \exp\left[-\frac{(\ell-\ell_{*})^2}{2s_{}^2}\right].
\end{align}
This integral can be computed exactly and leads to a result in terms of Hypergeometric function. We can arrive at the final answer also by using eq.~\eqref{def:B} and \eqref{eq:deform} with appropriate $h$ for the Gaussian case, which now reads for even $p$:
\begin{equation}
 \begin{aligned}
 z_{\mathtt{deformed}} &=
 \left( p! \sum_{r=p/2}^{p} \, \frac{1}{r!} B(r) \right)  z_\mathtt{SYK}= \frac{ 2^{p/2} s^p }{\sqrt{\pi}} \Gamma\left( \frac{1+p}{2} \right) {}_1F_1\left( -\frac{p}{2} , \frac{1}{2} , - \frac{\ell_*^2}{2s^2} \right)z_\mathtt{SYK}\label{1f1}, 
\end{aligned}
\end{equation}
Note that the sum runs from $p/2$ because  $B(r_1)=0$ for $r_1<p/2$. This happens because $\alpha_i$ can be $1$ or $2$ for the Gaussian deformation, leading to $2r\geq \sum_{i=1}^{r}\alpha_i=p$. Thus $\sum_{i=1}^{r}\alpha_i=p$  can not have solution if $r<p/2$. A similar calculation can be done for odd $p$ as well, where the sum runs from $(p+1)/2$, leading to a slightly different final expression. For simplicity, in what follows, we will assume $p$ is even without loosing any physical content.\\

In the two-time case, when $\mu =0$, one has:
$ \zeta(0)=  z^2_{\mathtt{SYK}} \prod_{i=L,R}\langle l_i^{p}\rangle_{s_i}$, which once again may either be computed exactly using eq.~\eqref{exact1} or using einbeins. When $\mu \neq 0$ we may write down the half-wormhole expansion as in eq.~\eqref{phik} : 
\begin{align}
\zeta_\mathtt{deformed}&= \sum_{k=0}^{p} \tilde \Phi_k \, \sum_{n=0}^{k} \frac{k!}{(k-n)! (nq)!} \left(\frac{q}{N}\right)^{n} \, \left(N \mu \right)^{nq} \prod_{i=L,R}\left( \int \mathrm{d} \ell_i  \, \ell_i^{p-n}\,e^{-S(\ell_i)} \right)\label{zeta-l*}
\end{align}
Since $S(\ell_i)$ is Gaussian the integral in parentheses can be performed exactly : 
\begin{equation}
\begin{aligned}
&\int_{-\infty}^{\infty } \mathrm{d} \ell  \, \ell^{p-n}\,e^{-S(\ell)} \\
&= \frac{ 2^{\frac{p-n}{2}} s^{p-n}}{\sqrt{\pi }}
\begin{cases} 
\Gamma \left(\frac{1}{2} (p-n+1)\right) \, _1F_1\left(\frac{n-p}{2};\frac{1}{2};-\frac{\ell_*^2}{2 s^2}\right)  & n\equiv p(\mathrm{mod} 2)\\
\sqrt{2} \frac{\ell_*}{s} \Gamma \left(\frac{p-n}{2} +1\right) \, _1F_1\left(\frac{1}{2} (n-p+1);\frac{3}{2};-\frac{\ell_*^2}{2 s^2}\right)& n+1\equiv p(\mathrm{mod} 2)\label{integral}
 \end{cases}
\end{aligned}
\end{equation}
In spite of the above answer for the integral, the finite sums in eq.~\eqref{zeta-l*} cannot be explicitly performed in closed form. However, in the small variance  $s\rightarrow 0$ coupled with the the large $p$ and $1/N \ll \mu \ll 1$ limit, the final term of the series eq.~\eqref{zeta-l*} (the wormhole saddle $= \vev{\zeta_\mathtt{deformed}}$) dominates and we find: 
\begin{align}
\zeta_\mathtt{deformed} &\underset{\underset{ p \gg 1}{1/N\ll  \mu \ll 1}} {\approx}\tilde\Phi_p \frac{\ell_*^{2p} }{q} \exp\left(\tfrac{\mu N}{l_*^{2/q}}\right).\label{limitzeta} 
\end{align}
Here we have used that in the small $s$ limit, the integral given by eq.~\eqref{integral} gives $\ell_*^{p-n}$.
This is consistent with the saddle point expectations in the zero variance limit. In this zero {\em uncertainty} regime of the einbeins, the deformation reduces to the scaling $H_0 \rightarrow \ell_* H_0$, which is same as $\bar{J} \rightarrow \ell_* \bar{J}$ and $\mu\to \mu/l_*^{2/q}$. This leads to the $ \ell_*^{2p}$ factor and rescaled $\mu$ in eq.~\eqref{limitzeta}.


\subsection*{Acknowledgements}

It is a pleasure to thank Ahmed Almheiri, Juan Maldacena, Dalimil Maz\'{a}\v{c}, and, Baur Mukhametzhanov. Both DD and AS would like to acknowledge the partial support provided by the Max Planck Partner Group grant MAXPLA/PHY/2018577. DD also  acknowledges partial support by the MATRICS grant {{SERB/PHY/2020334}}. SP acknowledges a debt of gratitude for the funding provided by Tomislav and Vesna Kundic as well as the support from the grant DE-SC0009988 from the U.S. Department of Energy. This work was performed in part at Aspen Center for Physics, which is supported by National Science Foundation grant PHY-1607611.


\appendix

\section{The $q$ dependence of the Scallop} 
\label{app:scallop}

\begin{figure}[!h] 
\centering
\includegraphics[scale=0.5]{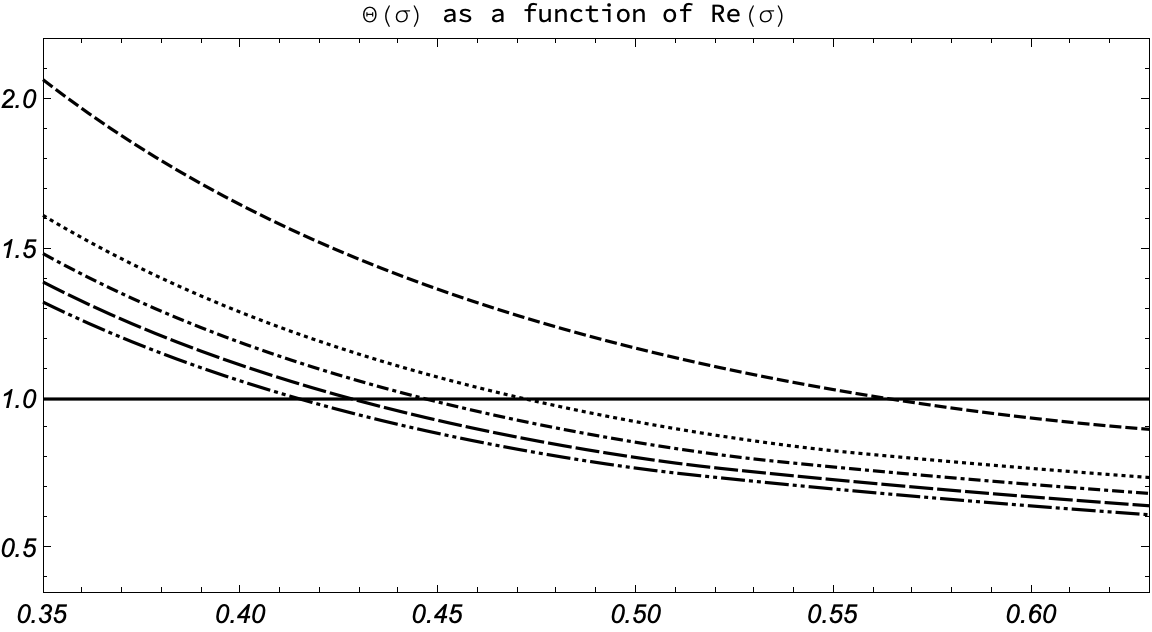} 
\caption{Here we have plotted $\Theta(\sigma)$ along the real $\sigma$ line for $q = 4,\, 8,\, 12,\, 20,\, 40 $. As $q$ increases, $\Theta(\sigma)$ becomes less than 1 for smaller values of $\sigma$, thus the scallop gets smaller.} \label{fig:scallop}
\end{figure}

Here we work with the undeformed theory and look into the $q$ dependence of the size of the scallop, ie, the half-wormhole dominated region which is by definition not self-averaging. To see this, one can calculate the ratio $\Theta(\sigma)$ defined in eq.~\eqref{Theta} on the complex $\sigma$ plane. If $\Theta(\sigma)>1$, then the half-wormholes dominate. Here we consider the case $\sigma_{LR'}\sigma_{RL'} = s^2 e^{2 \pi i m /q}$, $\sigma_{LL'}\sigma_{RR'} = 0$. For the ease of computation we also focus on $\Theta(\sigma)$ along the real $\sigma$ line, i.e.,\,\, for $m=0$. From Fig.\ref{fig:scallop}, we see that the scallop gets smaller as $q$ increases, thus the non self-averaging area on the complex $\sigma$ plane decreases for increasing $q$.

\section{A modified Bessel function identity }
\label{app:bessel}

We find for various values of $p$, the one-time prefactors of $z_\mathtt{SYK}$ from eq. \eqref{eq:deform} and eq. \eqref{eq:besselexact} match each other exactly. Below we show some numerical plots demonstrating this agreement.

\begin{figure}[htbp!]
 \centering
  \includegraphics[scale=0.7]{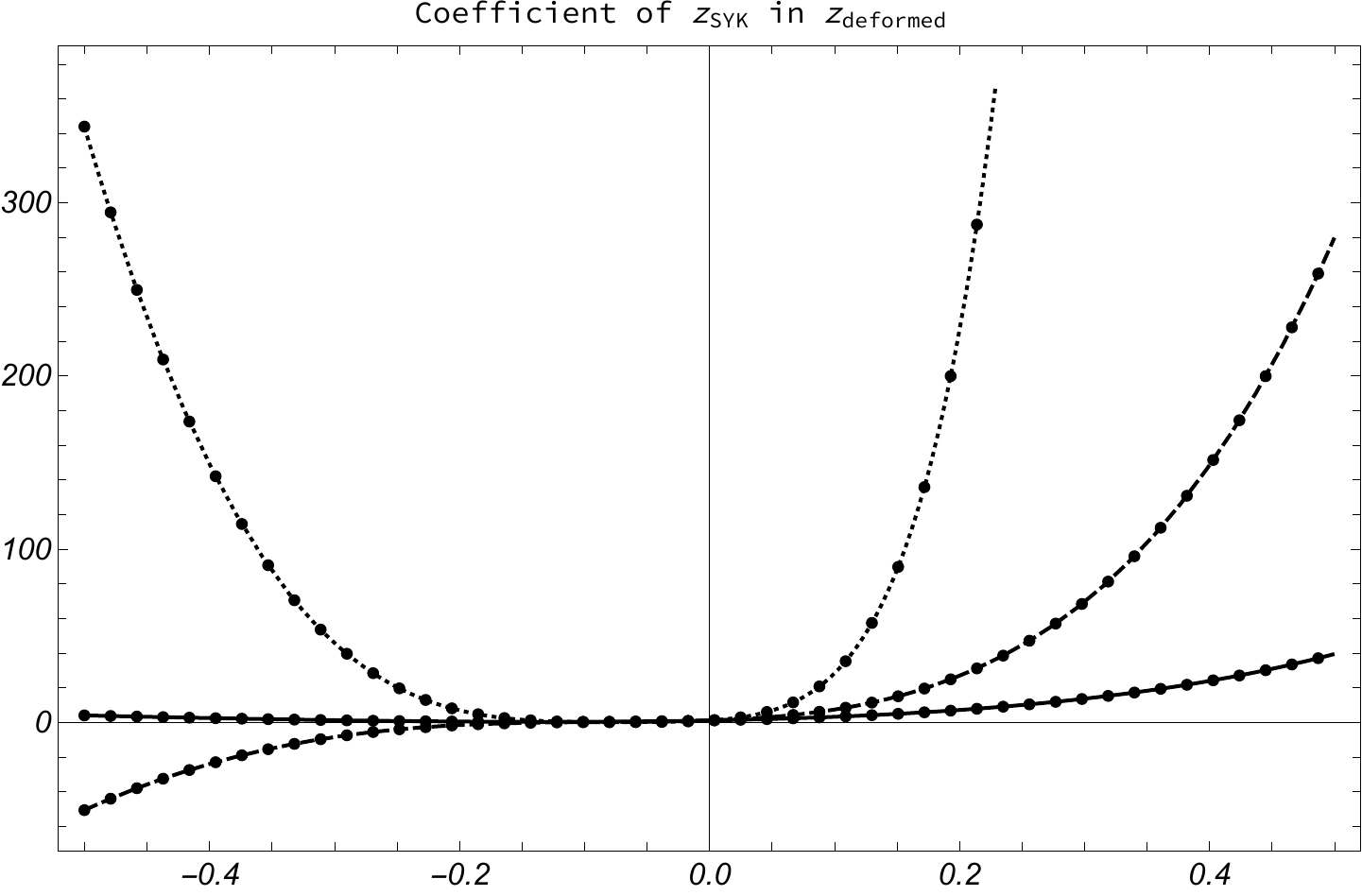} 
  \caption{The lines are for different values of $p$ and $E_0$ for the coefficients in eq.\eqref{eq:deform}. The black dots are corresponding points obtained using eq.\eqref{eq:besselexact}. Solid line is for $p=3, E_0 = -\tfrac{1}{16}$, dashed for $p=4, E_0=-\tfrac{1}{40}$ and dotted for $p=5, E_0 = -\tfrac{1}{15}$. }
  \label{fig:zcoeff}
 \end{figure}

If we set $\lambda=1$ and parametrize $x = \frac{\sqrt{1 + 8 E_0 }}{4}$ then the above match follows from the following series representation for modified Bessel $K$ with negative half integer index:
\begin{align}
K_{\tfrac{1}{2} - p }(x) &= \sqrt{2 \pi} e^{-x} p! \left( 4x \right)^{p-\tfrac{1}{2} } \sum_{r=1}^p \frac{(-x)^r}{r!} \sum_{\underset{\sum_{i=1}^{r}\alpha_i=p}{\{\alpha_i\geqslant 1\} }}\,\prod_{i=1}^{r}  \binom{\tfrac{1}{2}}{  \alpha_i }\left(-2 x^2 \right)^{-\alpha_i}\,,
\end{align}
which can be simplied into 
\begin{align}\label{eq:id}
K_{\tfrac{1}{2} - p }(x) &= \sqrt{\frac{2 \pi}{x}} e^{-x} p! 2^{p-1}\sum_{r=1}^p \frac{(-x)^{r-p}}{r!} \sum_{\underset{\sum_{i=1}^{r}\alpha_i=p}{\{\alpha_i\geqslant 1\} }}\,\prod_{i=1}^{r}  \binom{\tfrac{1}{2}}{  \alpha_i }\,.
\end{align}
We have checked  explicitly the expressions on both sides of the above equality for specific integer values of $p$ and found a match. In what follows, we sketch a proof of the identity.\\

$\bullet$ \textbf{Proof:} The identity given by eq.~\eqref{eq:id} can be proven in following manner. Let us focus on the r.h.s of \eqref{eq:id}, call it $I(x)$. Now we set $s=p-r$ to rewrite 
\begin{equation}
\begin{aligned}
I&= \sqrt{\frac{\pi}{2x}} e^{-x}\sum_{s=0}^{p-1} \frac{(-x)^{-s}}{(p-s)!}  p! 2^{p} \sum_{\underset{\sum_{i=1}^{p-s}\alpha_i=p}{\{\alpha_i\geqslant 1\} }}\,\prod_{i=1}^{p-s}  \binom{\tfrac{1}{2}}{  \alpha_i }\,,
\end{aligned}
\end{equation}
which can further be rewritten as
\begin{equation}
\begin{aligned}\label{eq:id2}
I&= \sqrt{\frac{\pi}{2x}} e^{-x}\sum_{s=0}^{p-1}\frac{p! s! }{(p-s)!}  \frac{(2x)^{-s}}{s!}  \left[  \sum_{\underset{\sum_{i=1}^{p-s}\alpha_i=p}{\{\alpha_i\geqslant 1\} }}\,\prod_{i=1}^{p-s} \frac{1}{\alpha_i} \binom{2\alpha_i-2}{  \alpha_i-1 }\right]\\
&= \sqrt{\frac{\pi}{2x}} e^{-x}\sum_{s=0}^{p-1}\frac{p! s! }{(p-s)!} \frac{(2x)^{-s}}{s!} \left[   \sum_{\underset{\sum_{i=1}^{p-s}\beta_i=s}{\{\beta_i\geqslant 0\} }}\,\prod_{i=1}^{p-s} \frac{1}{\beta_i+1} \binom{2\beta_i}{  \beta_i}\right]\\
&= \sqrt{\frac{\pi}{2x}} e^{-x}\sum_{s=0}^{p-1}\frac{p! s! }{(p-s)!} \frac{(2x)^{-s}}{s!} \left[   \sum_{\underset{\sum_{i=1}^{p-s}\beta_i=s}{\{\beta_i\geqslant 0\} }}\,\prod_{i=1}^{p-s} C_{\beta_i}\right]\,.
\end{aligned}
\end{equation}
Here $C_{\beta_i}$ is the $\beta_i$th Catalan number.
The generating function for the Catalan numbers is
\begin{equation}
\sum_{n=0}^{\infty}C_n x^n= \frac{1-\sqrt{1-4x}}{2x}\,.
\end{equation}
Using the generating function, we see that 
\begin{equation}
\begin{aligned}
\left[ \sum_{\underset{\sum_{i=1}^{p-s}\beta_i=s}{\{\beta_i\geqslant 0\} }}\,\prod_{i=1}^{p-s} C_{\beta_i}\right] &= \text{Coefficient of $x^{s}$ in}\ \left(\frac{1-\sqrt{1-4x}}{2x}\right)^{p-s}=\frac{p-s}{p+s}\binom{p+s}{s}
\end{aligned}
\end{equation}
A standard way to derive the above coefficient is to do a contour integral.

Now we plug the sum of product of Catalan numbers in eq.~\eqref{eq:id2} to obtain
\begin{equation}
\begin{aligned}
I&= \sqrt{\frac{\pi}{2x}} e^{-x}\sum_{s=0}^{p-1}\frac{p! s! }{(p-s)!} \frac{p-s}{p+s}\binom{p+s}{s} \frac{(2x)^{-s}}{s!} \\
&= \sqrt{\frac{\pi}{2x}} e^{-x}\sum_{s=0}^{p-1}\frac{(p+s-1)!}{(p-s-1)!} \frac{(2x)^{-s}}{s!} \\
&= K_{p-1/2}(x)=K_{1/2-p}(x)\,.
\end{aligned}
\end{equation}
This completes the proof. Note, here we have used $K_\nu(x)=K_{-\nu}(x)$ for $\nu \notin \mathbb{Z}$ and the following identity for Bessel function with half-integer index \cite{MR0010746} :
\begin{equation}
K_{n+1/2}(x)=\sqrt{\frac{\pi}{2x}} e^{-x}\sum_{s=0}^{n}\frac{(n+s)!}{(n-s)!} \frac{(2x)^{-s}}{s!}\,,
\end{equation} 
and set $n=p-1$.



\bibliographystyle{jhep}
\bibliography{ref-ttbsyk}

\end{document}